\documentstyle[epsf,epsfig,pre,aps,floats,twocolumn]{revtex}
\newcommand{\bm}[1]{\mbox{\boldmath $#1$}}
\newcommand{\R}{{\cal R}}
\begin{document} 
\title{Phase space embedding of electrocardiograms}

\author{Marcus Richter and Thomas Schreiber\\
Physics Department, University of Wuppertal, 42097 Wuppertal, Germany}

\maketitle
\begin{abstract} 
  We study properties of the human electrocardiogram under the working
  hypothesis that fluctuations beyond the regular structure of single cardiac
  cycles are unpredictable. On this background we discuss the possibility to
  use the phase space embedding method for this kind of signal. In particular,
  the specific nature of the stochastic or high dimensional component allows to
  use phase space embeddings for certain signal processing tasks. As practical
  applications, we discuss noise filtering, fetal ECG extraction, and the
  automatic detection of clinically relevant features. The main purpose of the
  paper is to connect results of embedding theory which had not been previously
  applied in practise, and practical applications which had not yet been
  justified theoretically.

\end{abstract}

\section{Introduction}
In nonlinear time series analysis, methods developed in nonlinear dynamics are
applied to time series data in order to capture as much of the underlying
structure as possible. One central question hereby is whether the dynamics of
the system can be reconstructed on the basis of the data given. Given the case
that the considered system is deterministic, the dynamics unambiguously evolves
one state in phase space into another. Therefore, the reconstruction of phase
space is a fundamental problem which plays the key role in many applications:
as long as the state space reconstruction remains unjustified all of the
consequent analysis may be wrong. From this point of view, we discuss the phase
space embedding of realistic signals in the example of the normal human
electrocardiogram. An introduction to nonlinear dynamics is found in
\cite{KaplanGlass,Ott}. An account of nonlinear time series methods is given by
\cite{KantzSchreiber}.

\begin{figure}
\centerline{\sf 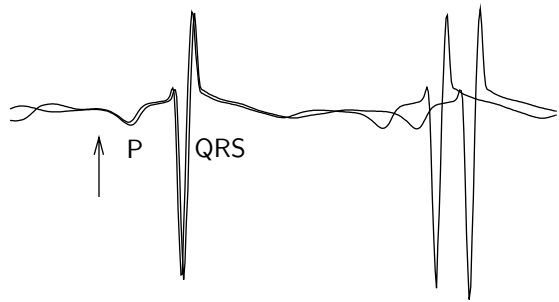}
\caption[]{Two different cycles of the same ECG have been aligned (indicated by
    the arrow) right before the beginning of the P-waves.  The stochastic
    components which lead to deviations from a pure limit cycle behavior
    manifest themselves in different cycle lengths and a slight variation of
    shape. \label{stoch_comp_FIG}}
\end{figure}

\begin{figure}
\centerline{\hspace*{0.2cm}\small \sf 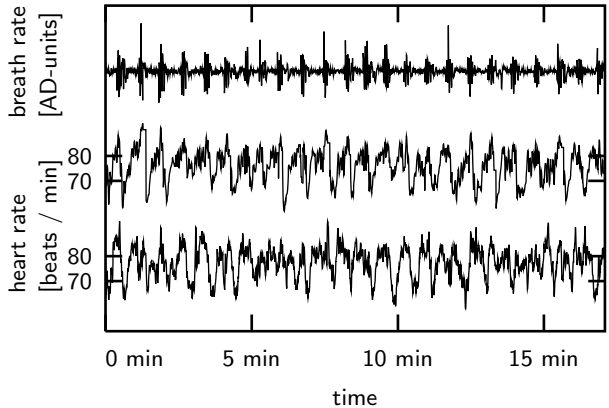}
\caption[]{Simultaneous measurements of breath and heart rates, upper and
    middle trace. The lower trace contains a randomized sequence preserving the
    autocorrelation structure and the cross correlation to the fixed breath
    rate series. The surrogate data mimic much of the structure contained in
    the data, but not all. \label{heartrate}}
\end{figure}

The electrocardiogram (ECG) is one of the most prominent clinical tools to
monitor the activity of the heart. In order to record an electrocardiographical
signal, metal electrodes are placed on the patient's chest wall and extremities
(for details see e.g.\ \cite{goldbook}). The potentials are generated by the
atrial and ventricular muscle fibers. Due to the placement of the electrodes on
the skin at some distance from the heart, the signals measured correspond to
action potentials which are averaged over large regions of tissue. The
spreading of the electrical activity over the cardiac muscle is controlled by
the conduction system of the heart.  Atrial and ventricular contraction and
relaxation, respectively, correspond to characteristic ECG-waves which are
traditionally labeled in alphabetic order beginning with the letter 'P'.  As
long as the average heart rate does not change dramatically, a non-pathological
ECG shows a nearly periodic structure which is due to the continuous generation
of action potentials and the fixed pattern according to which the electrical
activity spreads out over the cardiac muscle.  However, apart from the
deterministic structure some kind of variability can be found. On the one hand,
the length of the time interval between successive beats fluctuates -- to a
certain extend~-- around the mean heart rate. (In the medical literature the
inter beat-interval is often called RR--interval which is defined as the time
between two consecutive R--waves). On the other hand, a slight variation of the
cycle shape and amplitude can be observed. Both kinds of fluctuations can be
gathered from Fig.~\ref{stoch_comp_FIG} where two different cycles of the same
ECG sequence are shown. As one can see in Figure~\ref{heartrate}, part of the
variation of the heart rate (middle trace) of a human at rest can be connected
with the breath cycle (upper trace). (Data from the Santa Fe Institutte time
series contest \cite{gold}, see Ref.~\cite{sfibook}). The lower trace contains
a randomized sequence that has the same autocorrelation structure, and the same
cross correlation to the breath rate as the middle trace. For details as the
generation of these surrogate data, and further references see
Ref.~\cite{anneal}. This random surrogate data explain most but not all of the
underlying structure by linear correlations. One explanation for the remaining
structure might be the presence of a high dimensional or nonlinear stochastic
component. In this paper we adopt as a working hypothesis that the fluctuations
in the instantaneous heart rate are effectively unpredictable.

The Fourier domain is inappropriate to capture the structure described above,
since the variation in cycle length leads to a dominating broad band component.
An alternative would be to formulate the dynamics of the ECG cycle by a
stochastically driven model in a low dimensional phase space. Since this space
is not fully accessible by measurements, we propose to use the delay
reconstruction technique \cite{takens} as a convenient tool to reveal both the
regular and stochastic aspects of the electrocardiogram. Hereby, one attempts
to reconstruct the state variables $\bm \xi$ of the system represented in phase
space on the basis of the single lead measurement. Phase space reconstruction
has been developed for purely deterministic systems. We do not want to make
this assumption here. We will argue, however, why the technique can be
appropriate in this case despite the stochastic component of the signal.

\subsection{Delay reconstruction of electrocardiograms}
\begin{figure}
\centerline{\rule{1.7cm}{0cm} \sf 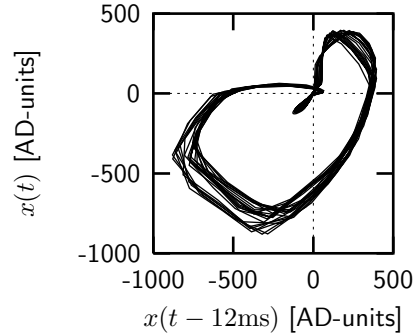}
\caption[]{Two dimensional delay reconstruction with delay time 12~ms using
    5000 data points of the whole ECG sequence from which two cycles are shown
    in Fig.~\ref{stoch_comp_FIG}. \label{delay_FIG}}
\end{figure}

In many cases of practical interest -- like in the ECG -- it is not possible to
measure the state variables of a system directly. Instead, the measuring
procedure yields some value $x= \varphi({\bm \xi})$, when the system is in
state $\bm \xi$. Here, $\varphi$ is a measurement function which in general
depends on the state variables in a nonlinear way. The time evolution of the
state of the system results in a scalar time series $x_1, x_2, x_3, \ldots \;
$. In order to reconstruct the underlying dynamics in phase space, delay
embedding techniques are commonly used. Delay vectors ${\bf x}_n = (x_n,
x_{n-l}, x_{n-2l}, \ldots, x_{n-(d-1)l})$ are a convenient method to transform
the scalar time series into $d$~dimensional vectors.  Herein, $d$ corresponds
to the embedding dimension while $l$ is the lag between the time series
elements.  In the case of purely deterministic systems, the embedding theorem
by Takens \cite{takens} and its generalization by Sauer {\it et al.}\ 
\cite{sauer} assure the equivalence between the reconstruction space and the
original phase space of the system under fairly broad conditions.
Figure~\ref{delay_FIG} shows a two dimensional reconstruction of the first 5000
time series elements (20s) of the ECG sequence underlying
Fig.~\ref{stoch_comp_FIG} by using delay coordinates at a lag of 12~ms.

Whilst for the signal processing applications discussed later usually much
higher embedding dimensions are needed, even the two dimensional representation
illustrates the underlying phase space structure succinctly. It is remarkable
that the signal structure in phase space is close to a manifold (which will be
exploited for noise reduction purposes discussed later). This suggests that the
underlying generating process is basically deterministic which manifests itself
in regular cycles, while the stochastic degrees of freedom lead to deviations
from the basic cycle. While the fluctuations in cycle shape are indicated in
the delay reconstruction, the stochastic variation of the time between cycles
cannot be seen directly. This can be easily understood in the context of the
following simplifying model. Let us suppose that the onset of a new cycle is
started by an external trigger signal and that the RR-interval time fluctuates
stochastically around the mean heart rate. Let us further assume that the ECG
signal rests on the base line after a cycle is finished, until a new trigger
signal starts the next depolarisation. Since the constant base line corresponds
to a single point in phase space, the stochastic variation of the cycle length
cannot be revealed using delay vectors, provided they span a time shorter than
the interbeat times. In order to reconstruct the underlying dynamics fully, it
is necessary to use some additional information -- like the trigger signal~--
describing the onset of the next cycle.

To conclude this introduction, the delay reconstruction technique can help to
reveal aspects of the nature of the underlying process, but it is not clear a
priori that the dynamics of the system can be reconstructed by using delay
coordinates in the case of stochastic forcing, even if the embedding dimension
is chosen sufficiently large. Fortunately, the embedding theorems proven
recently by Stark and coworkers \cite{stark,stark_proofs} (in the following we
refer to these theorems by the initials of the authors: SBDO) generalize the
well known Takens embedding theorem \cite{takens} and provide results for the
phase space reconstruction of stochastically forced dynamical systems. On the
basis of the SBDO-theorem it is possible to reconstruct the higher dimensional
phase space from the scalar time series data by using delay coordinates when
the sequence of stochastic influences that act on the system is known. Of
course, in the case of the ECG, we don't know the driving process. However, we
will argue that the necessary information can be recovered from the signal
a~posteriori.

\section{Embedding of stochastically driven signals}
In this section we give an overview over the fundamental notions introduced in
the papers by Stark and coworkers \cite{stark,stark_proofs} in order to state
the key points of the essential embedding theorem for stochastic systems. We
try to simplify the notation used as far as possible. The statement of the
theorem as well as the formalism introduced are illustrated in the examples
discussed subsequently.

In order to describe stochastic forcing, Stark considers skew product systems
(to avoid confusion we denote the state variables of the system by $\bm \xi$
while $\bf x$ labels the delay vectors build from the scalar time series
measurements in order to reconstruct the phase space of the system):
\begin{eqnarray*}
  {\bm \xi}_{n+1} & = & {\bf f}({\bm \xi}_n, \Omega_n) \\
  \Omega_{n+1} & = & \sigma(\Omega_n) \, .
\end{eqnarray*}
The dynamical system $\bf f$ is stochastically driven by the bi-infinite
sequence $\Omega = \ldots \omega_{-2} \omega_{-1} \omega_0 \omega_1 \omega_2
\ldots$, where $\Omega$ is build up by symbols $\omega_i$ that influence $\bf
f$. The time evolution of the stochastic sequence $\Omega$ is realized by the
map $\sigma$ shifting the elements of $\Omega$ to the left by one position. In
each time step, the dynamics $\bf f$ is affected only by the central element
$\omega_0$ of $\Omega$. It is natural to express the dependence of $\bf f$ on
$\omega_n$ by a sequence of maps ${\bf f}_n({\bm \xi})={\bf f}({\bm
  \xi},\omega_n)$, so that the dynamical system becomes:
\begin{equation}
   {\bm \xi}_{n+1} = {\bf f}_n ({\bm \xi}_n) \, .
\label{maps_eqn}
\end{equation}
In this framework it is possible to modify the underlying dynamics in each time
step. This formalism will be illuminated in an example given below where the
phase space reconstruction of a circular motion with fluctuating radius is
discussed.

The central theorem of \cite{stark_proofs} faces the question whether it
is possible to reconstruct the phase space of stochastically driven systems
only on the basis of the time series. In analogy to the usual embedding
techniques, one can define the $d$ dimensional delay embedding map $\Phi : M \to
{\R}^d$ for skew product systems by
\begin{equation}
\Phi_\omega({\bm \xi}_n) = \big( \, \varphi \left({\bm \xi}_n \right), \varphi
\left( {\bm \xi}_{n+1} \right), \ldots , \varphi \left( {\bm \xi}_{n+d-1} \right)
\! \big)
\label{embedding_frozen}
\end{equation}
(the usage of ``prelay'' rather than delay vectors is for technical reasons).
Here, the index $\omega$ has to be kept since the future of ${\bm \xi}_n$
depends explicitly on the stochastic sequence $\Omega$ under consideration.
This can be stressed by writing
\begin{eqnarray}
\Phi_\omega({\bm \xi}_n) & = & \big( \, \varphi({\bm \xi}_n), \varphi \left(
  {\bf f}_n ({\bm \xi}_n) \right), \ldots \nonumber \\
&& \ldots , \varphi \left( {\bf f}_{n+d-2} (\ldots {\bf f}_{n+1}({\bf f}_n ({\bm
  \xi}_n)) \ldots ) \right) \! \big)
\label{embedding_eqn}
\end{eqnarray}
where we have made use of eqn.~(\ref{maps_eqn}). Essentially, the main Theorem
(No 3.5 \cite{stark_proofs} and No 7 \cite{stark} respectively) says that if $d
\ge 2 d_0 +1$, then there is a {\it residual set} of dynamical systems $\bf f$ and
measurement functions $\varphi$, so that the delay embedding map $\Phi_\omega$
yields an embedding for almost every stochastic sequence
$\omega$. For the detailed genericity conditions on ${\bf f}$, $\varphi$ and
$\omega$ we refer to \cite{stark,stark_proofs}.

The crucial point here -- compared with the embedding theorems by Takens
\cite{takens} and Sauer {\it et al.} \cite{sauer} -- is that it is {\it not}
possible to reconstruct the underlying dynamics fully on the basis of only the
measured time series, even if the embedding dimension $d$ is sufficiently large
and the dynamical system $\bf f$ as well as the measurement function $\varphi$
are generic. In order to take the stochastic nature of the process into
account, it is necessary to use additional information describing the
stochastic influences. To be precise, the knowledge of the underlying
stochastic sequence is essential for the definition of the delay embedding map
(\ref{embedding_eqn}).

As an illustrative example consider the map
\begin{eqnarray}
\underbrace{\left( \begin{array}{c} x_{n+1} \\ y_{n+1} \end{array}
  \right)}_{\bm \xi_{n+1}}
& = &
\underbrace{\frac{r_{n+1}}{r_{n}} {\bf R}_{[\phi]}}_{{\bf f}_{n}}
\underbrace{\left(\begin{array}{c} x_{n} \\ y_{n} \end{array} \right)}_{{\bm
  \xi}_n} \label{circfluct_eqn} \\ \nonumber \\
& = & r_{n+1} {\bf R}_{[(n+1)\phi]} \left( \begin{array}{c} x_0 \\ y_0
  \end{array} \right) \nonumber \end{eqnarray}
where $ {\bf R}_\phi $ is a rotation by $\phi$ 
\[ {\bf R}_\phi = \left( \begin{array}{lr} \cos(\phi) & -\sin(\phi) \\ 
\sin(\phi) & \cos(\phi) \end{array} \right) \]
and the radius $ r_{n} $ is chosen at random between $r_{\rm min}$ and
$r_{\rm max}$ while the rotation angle $ \phi $ is fixed and generically
incommensurate with $2 \pi$.

\begin{figure}
\centerline{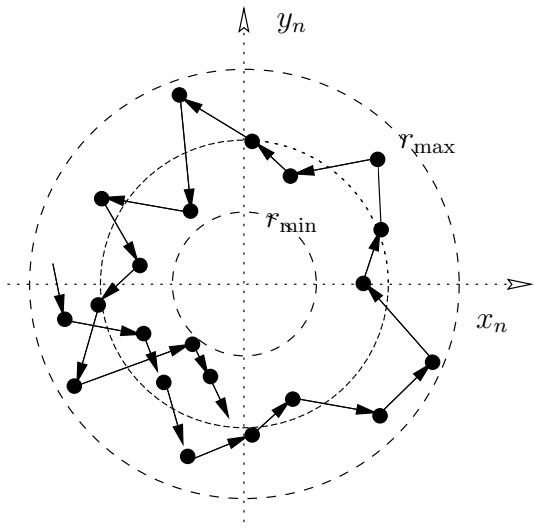}
\caption[]{Illustration of the phase space structure underlying equation
    (\ref{circfluct_eqn}). After many iterations, the dots evenly fill the disk
    area between $r_{\rm min}$ and $r_{\rm max}$. \label{circfluct_fig}}
\end{figure}

For simplicity, let the measurement function $\varphi$ ``measure'' the first
coordinate:
\[
  \varphi({\bm \xi}_n) = x_n \; .
\]
The central question is now whether the scalar time series $ x_0, \ldots, x_n,
x_{n+1}, \ldots $ yields an appropriate basis to reconstruct the dynamics of
the system. Fig.~\ref{circfluct_fig} contains a schematic representation of the
phase space structure underlying equation (\ref{circfluct_eqn}). Since the
fluctuations of the radius are bounded by $r_{\rm min}$ and $r_{\rm max}$, the
points generated according to Eq.~(\ref{circfluct_eqn}) are sprinkled over a
disk. This disk corresponds to the compact manifold $M$ in the theorems by
Stark {\it et al.}

For a moment consider the case that the sequence of radii would be fixed to
one: $r_n \equiv 1$. In this specific case it is obvious that the two
dimensional delay reconstruction map $\Phi({\bf x}_{n-1}) = (x_{n-1}, x_n)$ --
which is based on the measurement function specified above -- yields an
embedding for almost every value of $\phi$. This is simply due to the fact that
a uniform motion on a circle can easily be reconstructed by its first Cartesian
coordinate since both coordinates are related by a constant phase shift. As far
as a {\it deterministic} variation of the radius is involved, -- according to
Takens' theorem -- at least three dimensional delay embedding maps are needed
to avoid self intersections in reconstruction space. It can be seen quickly
that a pure delay embedding technique in which the dimension takes into account
only the degrees of freedom of $\bf f$ is in general not sufficient to
reconstruct the underlying dynamics in the case of {\it stochastic forcing}. In
contrast to deterministic systems, due to the stochastic influences crossing
points can appear between the trajectories on the manifold $M$ as indicated in
Fig.~\ref{circfluct_fig}. In order to determine the continuation of the
trajectories, it is necessary to use some information beyond the delay
coordinates. Fortunately, as Stark points out in the theorem cited above, the
additional knowledge of the driving stochastic sequence allows the full
reconstruction of the underlying dynamics.  Referring to map
(\ref{circfluct_eqn}), the knowledge of the sequence of radii $r_n$ allows the
continuation of the trajectories beyond the crossing points.

Apart from the pure illustrating function regarding the basic idea underlying
Stark's theorem, the model discussed can help to motivate the embedding of more
realistic signals like ECG sequences. As indicated in fig.~\ref{delay_FIG}, the
delay reconstruction of a normal human electrocardiogram bears resemblance to a
limit cycle structure with fluctuations. As pointed out in the introduction,
the fluctuations affect both the cycle length and its shape. One natural way to
describe the stochastic influences would be by trying to observe the driving
input signal. However, the variable parameters that influence the ECG waveform
are not usually accessible to measurements and in fact, are not fully known.
This is a general problem with the approach behind the work of Stark {\it et
  al.}, as well as that of Casdagli~\cite{Casdagli}: regarding a system as an
input-output device is only useful for the analysis of time series if both, the
output and the input sequences are observed.

\begin{figure}
\centerline{\hspace{2cm}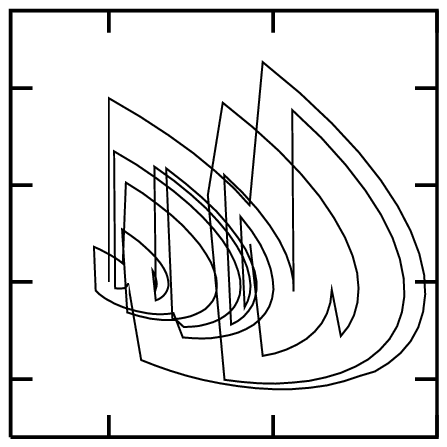}
\centerline{\hspace{2cm}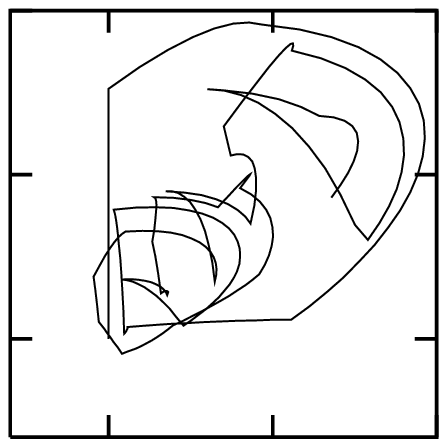}
\caption[]{Trajectories of a kicked, damped harmonic oscillator. Upper: true
    phase space. Lower: delay embedding. Kicks occur close in
    time.\label{fig:kriech1}}
\end{figure}
 
One of the main ideas we want to put forth in this paper is that there are
situations in which the input sequence can to some extent be inferred from the
observed data. In such a case, the embedding procedure may be applied
successfully, on the theoretical foundation of the theorem discussed above.  In
the case of the ECG, the stochastic driving mainly affects the duration of the
cycles, which is accessible a posteriori by measuring the RR intervals.  If
there is a variation of the ECG waveform itself which can be captured by a few
parameters, these can also be measured once the ECG has been recorded.  The
fact we use in doing this is that ECG recordings contain considerable
redundancy. With regard to ECG signals this implies that the explicit
specification of the underlying sequence $\Omega$ is not strictly necessary to
make use of the embedding technique.

\begin{figure}[t]
\centerline{\hspace{2cm}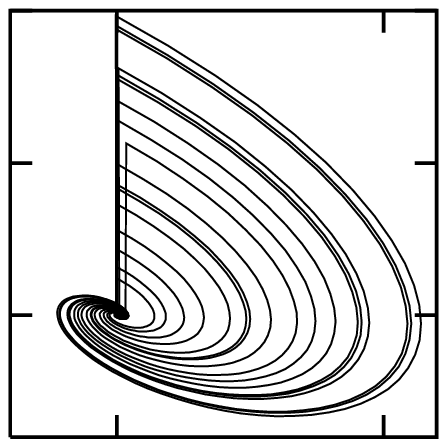}
\centerline{\hspace{2cm}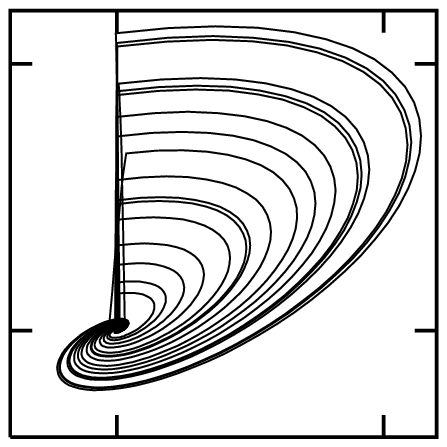}
\caption[]{Same as Fig.~\ref{fig:kriech1}, but kicks are well separated in time
    such that the system can relax between kicks.\label{fig:kriech2}}
\end{figure}
 
Let us study one more toy model to illustrate how the lacking
information about the input sequence may be contained in a signal from
an input-output system. Consider a damped harmonic oscillator
\begin{equation}
   \ddot{x} + \dot{x} + x = a(t)
\,,\end{equation}
where the driving term is zero except for kicks of random strength at times
$t_i$ such that the inter-beat intervals, $t_i-t_{i-1}$ are random in the
interval $[p,q]$. 
Figures~\ref{fig:kriech1} and~\ref{fig:kriech2} show two trajectories of such a
system with different choices of the inter-beat time interval $[p,q]$.  On
discretisation, the kicks are realized by finite jumps by a random amount in
the interval $[0,1]$.  The upper panel in each of these figures shows the true
phase space, while in the lower panels a delay representation is used with a
delay of one time unit. In Fig.~\ref{fig:kriech1}, beats are initiated with a
time separation of $p=0$ to $q=T/2$ where $T=4\pi/\sqrt{3}\approx 7.26$~time
units is the period of oscillation. This does not allow the system to relax
sufficiently between kicks in order to form characteristic structure in phase
space. Consequently, an embedding provides no clear picture. In
Fig.~\ref{fig:kriech2}, no kicks were closer in time than $p=T$, the maximal
separation being $q=3T$. The inter-beat parts of the trajectory are distinct
because of the different kick strength, but since this is the only fluctuating
parameter, they are essentially restricted to a two-dimensional manifold. This
manifold is preserved under time delay embedding although neither the sequence
of beat times nor the beat amplitudes are used explicitly. The randomness
prevails in the indeterminacy at the origin as to when the next beat will
occur.

\section{Predicting and projecting ECG signals}
\begin{figure}
\centerline{\sf 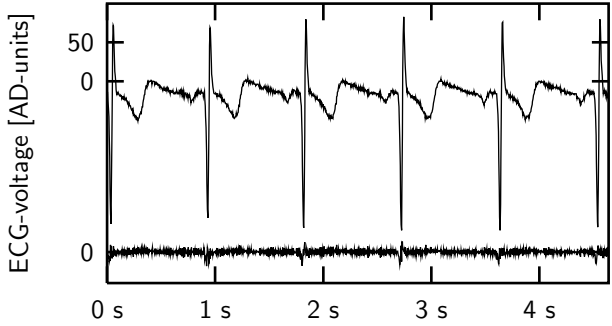} 
\caption[]{Deviations between the ECG sequence \cite{MIT} (upper trace) and one
    step predictions generated according to Eq.~(\ref{radialbasis_eqn}). The
    signals are measured in uncalibrated AD-units.\label{onestep_FIG}}
\end{figure}

While the structure of a normal human ECG is well predictable on a scale
considerably shorter than one cycle, fluctuations in cycle length limit the
accuracy at the onset of the QRS--complexes. Therefore, large deviations can be
found between the original ECG data and the one step predictions generated as
soon as a new QRS--complex appears, whilst the difference between both signals
is usually comparable to the magnitude of the noise level.
Fig.~\ref{onestep_FIG} illustrates these deviations with the ECG sequence
plotted. The ansatz chosen makes explicit use of the phase space reconstruction
by delay vectors ${\bf x}_n$. The ECG data considered are highly oversampled so
that $x_{n+1}$ is close to $x_n$ up to some corrections which are modeled by
radial basis functions:
\begin{equation}
x_{n+1} = \underbrace{x_n + \sum_{k=1}^K a_k \exp\left(-b\left[{\bf x}_n
      - {\bf c}_k\right]^2\right) + C}_{F({\bf x}_n)} \, .
\label{radialbasis_eqn}
\end{equation}
The constant $C$ allows to handle a zero offset directly instead by fitting via
radial basis functions, which would affect the quality of the parameters $a_k,
b$ and $c_k$. In order to place the centers $c_k$ of the basis functions, a
grid-based algorithm was used. The result plotted in Fig.~\ref{onestep_FIG} was
achieved by aspiring a uniform density of centers over the phase space regions
occupied by the reconstruction vectors and a total number of 10 basis
functions. Increasing the number of basis functions leads to a better
approximation of the ECG signal and therefore a lower in-sample prediction
error. In Fig.~\ref{onestep_FIG} the prediction error during the interbeat
times is comparable to the magnitude of the original ECG data noise level. It
is remarkable that the error of the one-step predictions is much larger at the
onset of the QRS-complexes so that the overall distribution is nonuniform.
Therefore one might conclude that the onset of the next QRS-complex remains
unpredictable.

To reduce the uncertainty underlying the onset of the next QRS--complex one
could think of using the length of the following RR--interval as an additional
input signal. It might be possible to lower the large prediction error at the
beginning R--wave by this technique, since the moment of the spike-like
deflection is now completely determined. The use of an additional input
sequence which characterizes the stochastic degrees of freedom corresponds to
the fixing of the stochastic sequence $\Omega$ in the formalism introduced by
Stark. Since the prediction of ECG signals is not in itself of much practical
relevance, we do not describe further details. The concrete implementation is
technically involved and would go far beyond the scope of this paper.

\begin{figure}
\centerline{\hspace*{-1cm} \epsfxsize=5cm \epsfbox{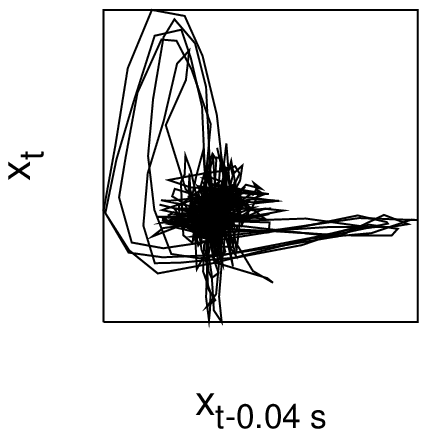} \hspace*{-1cm}
  \epsfxsize=5cm \epsfbox{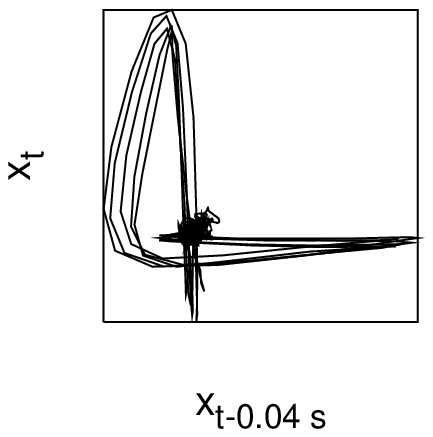} \vspace*{5mm}}
\caption[]{Delay reconstruction of the abdominal ECG recording (left panel) and
    the maternal component (right panel).\label{fecg_del}}
\end{figure}

\begin{figure}
\centerline{\footnotesize 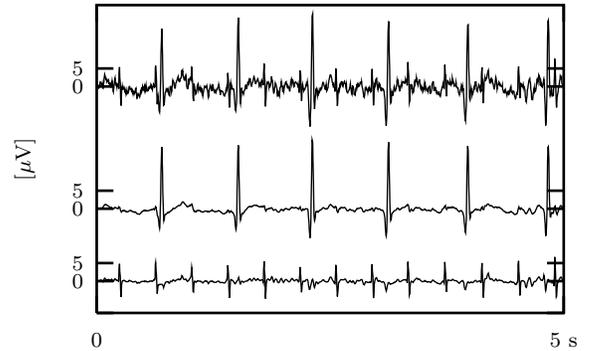}
\caption[]{Fetal ECG extraction by locally linear projections in phase
    space. The abdominal recording (upper trace) contains a dominant maternal
    component and the fetal signal which is covered by noise. The first noise
    reduction step yields the reconstructed maternal signal (middle) which
    allows to separate the noisy fetal ECG from the original recording. A
    second projection can be applied to clean up the fetal component (lower
    trace).\label{fecg}}
\end{figure}

State space reconstruction techniques form an essential tool for noise
reduction purposes. In order to separate the signal from its noise components,
basically two different techniques can be pursued. (I) On the one hand a global
fit to the dynamical equations can be used for the reconstruction of a
noiseless trajectory of the system, as it has been done in \cite{davies}, for
example. In this context, state space reconstruction techniques are involved to
capture the underlying dynamics. For example in Eq.~(\ref{radialbasis_eqn}) the
reconstruction technique enters via the delay vectors ${\bf x}_n$. As an
iterative method for reaching a trajectory which obeys the reconstructed
dynamics, gradient descent methods have been investigated in \cite{davies}.
(II) On the other hand it is possible to exploit the local structure in the
reconstruction space directly for noise reduction purposes without referring to
the global underlying dynamics. Methods based on this local projective approach
have been proposed in \cite{on}, see \cite{ks} for a review. The idea is that
if it is found empirically that the data points in reconstruction space are
located close to a manifold, the error by projecting onto that manifold may be
smaller than the error due to the noise.  If the trajectory of the system lies
on a low dimensional attractor, the projection technique can indeed be shown to
reduce noise. But also for non-deterministic signals, nonlinear noise reduction
techniques have been successfully applied for signal processing tasks like ECG
noise reduction \cite{SK,processing} or fetal ECG extraction \cite{SK2,marcus}.
The basic reason why this works is the mechanism illustrated in
fig.~\ref{fig:kriech2}. These signals, in spite of being non-deterministic,
nevertheless are near a low-dimensional manifold. Let us give the extraction of
the fetal ECG as an application of the delay embedding of ECG signals (fetal
electrocardiography is the only method to monitor the cardiac activity of the
fetus in a non invasive way). Figure~\ref{fecg_del} contains a two dimensional
delay representation of an abdominal ECG recording (left panel) and the
reconstructed maternal manifold structure (right panel) by the projective noise
reduction approach. Segments of the corresponding time series are shown in
Fig.~\ref{fecg} (data by courtesy of J.~F.~Hofmeister \cite{recording}). The
upper trace contains the abdominal signal (electrodes have been placed on the
maternal abdomen to record most of the fetal heart activity) while the middle
trace shows the abdominal projection of the maternal electrocardiogram which
has been extracted by noise reduction. The difference between the upper and
middle trace yields the noisy fetal component, which can be cleaned up in a
second noise reduction step (lower trace). The extraction of the fetal
electrocardiogram is broadly described in \cite{IEEE}, for a powerfull
modification of the projective algorithm which allows the real time extraction
of the fetal electrocardiogram on a Laptop PC (Pentium processor at 133 MHz,
Linux operating system, 250 Hz sampling rate) see \cite{Filter}.

As another practical application, phase space reconstruction techniques can be
employed for the localization of fundamental ECG waves like P-, R- or T-waves.
Instead of comparing the structures directly on the basis of the time series,
similar ECG waves can be identified as adjacent curves in reconstruction space.
For example the smaller loop to the lower left side of the origin in
fig.~\ref{delay_FIG} corresponds to the P-waves of the full ECG sequence
partially plotted in fig.~\ref{stoch_comp_FIG}, while the larger framing loop
is due to the QRS-complex. For the automatic detection of ECG cycle
substructures, delay vectors are defined to be neighbours if their distance to
the vector under consideration is smaller than a given size (for neighbour
searching methods cf.~e.g.\ \cite{KantzSchreiber}). Nearby trajectories can be
identified in phase space by searching for neighbours of the reconstruction
vectors of a specified structure. On the basis of the original time series,
these trajectories correspond to segments which are similar to the specified
one. By tagging a P-, R- or T-wave of a given electrocardiogram, previous or
following waves of this type can be located automatically.

To conclude, embedding techniques form a fundamental starting point for many
applications. Therefore, it is desirable to motivate the reconstructability of
the phase space in the case of typical time series. We illuminated this issue
by considering the example of the human electrocardiogram. Under the working
hypothesis that fluctuations beyond the regular structure of the cardiac cycle
are unpredictable we found the embedding theorems of Stark and coworkers
\cite{stark,stark_proofs} to be useful to motivate the embedding of ECG data.
Finally, we discussed useful applications which are based on phase space
reconstructions of the electrocardiogram in order to give examples of how
embedding techniques can be employed for practical tasks.

\section*{Acknowledgements}
We thank John F. Hofmeister and Petr Saparin for providing ECG recordings,
Holger Kantz and Rainer Hegger for stimulating discussions.  This work was
partially supported by the SFB 237 of the Deutsche Forschungsgemeinschaft.

\end{document}